\documentclass[aps,prd,twocolumn,superscriptaddress,nofootinbib,showpacs]{revtex4-1}

\usepackage{amsmath,amssymb}
\usepackage{graphicx}

\begin{document}

\title{Neutrino masses and mixing in $A_4$ models with three Higgs doublets}

\author{R. Gonz\'{a}lez Felipe}\thanks{E-mail: ricardo.felipe@ist.utl.pt}
\affiliation{Instituto Superior de Engenharia de Lisboa - ISEL,
	1959-007 Lisboa, Portugal}
\affiliation{Centro de F\'{\i}sica Te\'{o}rica de Part\'{\i}culas (CFTP),
    Instituto Superior T\'{e}cnico, Universidade T\'{e}cnica de Lisboa,
    1049-001 Lisboa, Portugal}
\author{H.~Ser\^{o}dio}\thanks{E-mail: hugo.serodio@ific.uv.es}
\affiliation{Departament de F\'{\i}sica Te\`{o}rica and IFIC,
		Universitat de Val\`{e}ncia-CSIC,
		E-46100, Burjassot, Spain}
\author{Jo\~{a}o P.~Silva}\thanks{E-mail: jpsilva@cftp.ist.utl.pt}
\affiliation{Instituto Superior de Engenharia de Lisboa - ISEL,
	1959-007 Lisboa, Portugal}
\affiliation{Centro de F\'{\i}sica Te\'{o}rica de Part\'{\i}culas (CFTP),
    Instituto Superior T\'{e}cnico, Universidade T\'{e}cnica de Lisboa,
    1049-001 Lisboa, Portugal}

%\date{\today}

\begin{abstract}
We study neutrino masses and mixing in the context of flavor models with
$A_4$ symmetry, three scalar doublets in the triplet representation, and
three lepton families. We show that there is no representation assignment
that yields a dimension-5 mass operator consistent with experiment. We
then consider a type-I seesaw with three heavy right-handed neutrinos,
explaining in detail why it fails, and allowing us to show that agreement
with the present neutrino oscillation data can be recovered with the
inclusion of dimension-3 heavy neutrino mass terms that break softly
the $A_4$ symmetry.
\end{abstract}

\pacs{12.60.Fr, 14.80.Ec, 11.30.Qc, 11.30.Ly}

\maketitle

\section{Introduction}
\label{sec:intro}

One can attempt to explain the structure of neutrino masses and mixing by
building full theories subject to discrete symmetries (for recent reviews, see
e.g. Refs.~\cite{Altarelli:2010gt,Ishimori:2010au,Grimus:2011fk,King:2013eh}).
As an alternative, one may consider effective operators in the low-energy
limit and their remnant symmetries, without recourse to the full
theory~\cite{dois}. In either case, it is important to ensure that the vacuum
structure chosen for the scalar sector of the theory does indeed correspond to
a global (and not merely a local) minimum. Recently~\cite{ivanov_min}, the
global minima of theories with three Higgs doublets, $\Phi_k$ ($k=1,2,3$), in
triplet representations of $A_4$~\cite{merlo} or $S_4$ have been identified.
It turns out that the allowed alignments for the vacuum expectation values
(VEVs) in $A_4$ are~\cite{ivanov_min}
\begin{eqnarray}\label{vev:A4}
&&v\, (1,0,0),\nonumber\\
&&v\, (1,1,1),\\
&&v\, (\pm 1, \eta, \eta^*)\ \textrm{with }\ \eta = e^{i \pi/3},\nonumber\\
&&v\, (1, e^{i \alpha}, 0)\ \textrm{with any phase}\ \alpha.\nonumber
\end{eqnarray}
Permutations of these VEVs are still global minima; other solutions of the
stationarity conditions are not. The quark sector of such theories was
explored by us in Ref.~\cite{nos}. It was shown that, at tree level, there is
no consistent theory with only three families of standard quark fields that
can explain the fact that the quark masses and the Cabibbo-Kobayashi-Maskawa
CP-violating phase are nonvanishing.

In this article, we extend the analysis of Ref.~\cite{nos} to the neutrino
sector, assuming that neutrino masses are generated through a low-energy mass
operator in an effective theory. We consider models with three Higgs doublets
$\Phi_k$ in a triplet representation of $A_4$, so that the only possible VEV
structures are those in Eq.~\eqref{vev:A4}.

In Sec.~\ref{sec:A4lepton} we recall the features that apply to the
charged-lepton Yukawa matrices in $A_4$~\cite{nos}. Then, in
Sec.~\ref{sec:A4neutrino}, we turn to the neutrino sector. In
Sec.~\ref{sec:effective}, we address the question of whether it is possible
to get a consistent picture in an effective theory with the following
particle content: three scalar doublets in a triplet representation of $A_4$,
three left-handed lepton doublets $L_L$ and three right-handed charged
leptons $\ell_R$ in any representation of $A_4$. We conclude that, although
an $A_4$-invariant dimension-5 effective operator $( L_L \Phi)(L_L \Phi)$ can
always be built, it is not possible to assign suitable $A_4$ representations
to the fields in order to obtain viable (nondegenerate and nonvanishing) mass
spectra, with the VEV alignments given in Eq.~\eqref{vev:A4} and neutrino
masses arising solely from the effective operator. As a result, in such a
minimal $A_4$ framework, none of the standard seesaw mechanisms, usually
invoked to give small masses to neutrinos, is consistent with the
experimental data. With the aim of identifying what features need to be
corrected when building a complete viable model, in Sec.~\ref{sec:IandIII} we
discuss in detail type-I (type-III) seesaw models~\footnote{We do not
consider here the type-II seesaw scenario since it involves triplet scalars,
thus making, in general, the identification of the global minima of the
scalar potential extremely hard.} that contain a minimal particle content,
namely, three generations of left-handed lepton doublets and right-handed
charged lepton singlets, and three right-handed neutrino singlets $\nu_R$
(fermion triplets $\Sigma_R$). Since, at the Lagrangian level, the $A_4$
flavor group structure is the same for both seesaw cases, so are the
conclusions. Clearly, the above minimal setup is not sufficient to build a
consistent $A_4$ flavor model that leads to nonzero nondegenerate
charged lepton and neutrino masses, and to the correct leptonic mixing. The
detailed discussion presented in Sec.~\ref{sec:IandIII} is used in
Sec.~\ref{sec:soft} to uncover examples where a soft breaking of the $A_4$
symmetry is enough to provide consistency with experiment. Our conclusions
are briefly summarized in Sec.~\ref{sec:conclusions}.

\section{$A_4$ and charged leptons}
\label{sec:A4lepton}

The $A_4$ group has four irreducible representations: three singlets $\
\mathbf{1}, \ \mathbf{1^{\prime}}, \ \mathbf{1^{\prime \prime}}$, and one
triplet $\mathbf{3}$. Here we adopt the basis used in Ref.~\cite{nos} for the
generators of the group, and place the three Higgs doublets in the triplet
representation, $\Phi \sim \mathbf{3}$. The conclusions reached in Table~II
of Ref.~\cite{nos} for the down-type quarks also hold for charged leptons. In
particular, requiring nonvanishing nondegenerate charged lepton masses forces
the VEV alignments to be $v(1,1,1)$ or $v(\pm1, \eta, \eta^\ast)$,
restricting the possible representation assignments to the cases listed in
Table~\ref{tabA4:charged}.
\begin{table}[h!]
\begin{center}
\begin{tabular}{ccccc}
\hline\hline
& $L_L$ & & $\ell_R$ & \\
\hline
& $\mathbf{3}$ & & $\mathbf{3}$ & \\
& $\mathbf{3}$ & & $(\mathbf{1}, \mathbf{1^\prime}, \mathbf{1^{\prime \prime}})$ & \\
& $(\mathbf{1}, \mathbf{1^\prime}, \mathbf{1^{\prime \prime}})$ & & $\mathbf{3}$ & \\
\hline\hline
\end{tabular}
\end{center}
\caption{Possible representations of the left-handed lepton doublet
($L_L$) and right-handed charged-lepton singlets ($\ell_R$), which lead
to nonvanishing nondegenerate charged lepton masses, when the three Higgs
doublets are in a triplet representation $\mathbf{3}$. The case $L_L \sim
\mathbf{3}$, $\ell_R \sim \mathbf{3}$ leads to nonrealistic
charged-lepton masses. \label{tabA4:charged}}
\end{table}

Let us consider in detail the case $L_L \sim \mathbf{3}$, $\ell_R \sim
\mathbf{3}$. We may parametrize the charged lepton mass matrix
as
\begin{equation}
m_\ell =
\left(
\begin{array}{ccc}
0 & b\, e^{i \beta}\, v_3 & a\, e^{i \alpha}\, v_2\\
a\, e^{i \alpha}\, v_3 & 0 & b\, e^{i \beta}\, v_1\\
b\, e^{i \beta}\, v_2 & a\, e^{i \alpha}\, v_1  & 0
\end{array}
\right).
\label{Ll3lR3}
\end{equation}
Taking the VEV alignment $(1,1,1)$ or $(\pm 1, \eta,
\eta^*)$,\footnote{Henceforth, we assume without loss of generality that
$v=1$.} the eigenvalues of $m_\ell m_\ell^\dagger$ are
\begin{eqnarray}
& & a^2 + b^2 + 2 a b \cos{(\alpha-\beta)},\nonumber\\
& & a^2 + b^2 - 2 a b \cos{(\alpha-\beta \pm \pi/3)},
\label{33charged}
\end{eqnarray}
which coincide with the squared masses of the charged leptons. Although the
eigenvalues depend on three parameters, one can see that they cannot be used
to fit the hierarchical structure of the charged lepton masses. Indeed, it is
easy to show that
\begin{eqnarray}
&&
3 (a^2 + b^2)
=
m_e^2 + m_\mu^2 + m_\tau^2\, ,
\\
&&
3 a b
=
\sqrt{m_e^4 + m_\mu^4 + m_\tau^4 - m_e^2 m_\mu^2 - m_e^2 m_\tau^2 - m_\mu^2 m_\tau^2}\, .
\nonumber
\end{eqnarray}
Substituting for the experimental masses~\cite{pdg}, we find that $(a - b)^2
\simeq -1.05 \textrm{ GeV}^2$. Thus, in this case, the charged lepton masses
cannot be fitted.

\section{$A_4$ and Majorana neutrinos}
\label{sec:A4neutrino}

If neutrinos were Dirac particles, with masses arising from Yukawa couplings
with $\Phi_k$, our conclusions would be the same as those reached in
Ref.~\cite{nos} for the up-type quarks. Clearly, in that case, the existence
of one massless neutrino or the lack of leptonic CP violation would not
contradict current experiments~\cite{Branco:2011zb}. In this work, however,
we focus on Majorana neutrinos whose masses are given by some seesaw
mechanism or else are generated effectively by integrating out unspecified
heavy degrees of freedom.

\subsection{Effective dimension-5 operator for neutrino masses}
\label{sec:effective}

There is only one dimension-5 operator made out of the SM fields which
respects the SM gauge symmetry~\cite{weinberg},
\begin{equation}
{\cal L}_{\textrm{eff}}
=
\left( \bar{L}_L \tilde{\Phi} \right)\,
K\,
\left( \tilde{\Phi}^T L_L^c \right)
+ \textrm{H.c.},
\label{L_eff}
\end{equation}
where $K$ is a complex symmetric matrix, and $\tilde{\Phi} = i \sigma_2
\Phi$. Since we are only interested in the group structure, we will only
refer to $L_L L_L$ and $\tilde{\Phi} \tilde{\Phi}$.

If $L_L \sim \mathbf{3}$, we can form the symmetric bilinears
\begin{eqnarray}
(L_L \otimes L_L)_{\mathbf{1}}
&=&
L_1^2 + L_2^2 + L_3^2,
\label{33_1}
\\
(L_L \otimes L_L)_{\mathbf{1^\prime}}
&=&
L_1^2 + \omega^2 L_2^2 + \omega L_3^2,
\label{33_1p}
\\
(L_L \otimes L_L)_{\mathbf{1^{\prime \prime}}}
&=&
L_1^2 + \omega L_2^2 + \omega^2 L_3^2,
\label{33_1pp}
\\
(L_L \otimes L_L)_{\mathbf{3}s}
&=&
2 (L_2 L_3, L_3 L_1, L_1 L_2),
\end{eqnarray}
where $\omega = e^{2 i \pi/3}$. Similar combinations hold for $\tilde{\Phi}
\tilde{\Phi}$. Forming all combinations that can lead to a singlet, we find
\begin{eqnarray}
{\cal L}_{\textrm{eff}}
&=&
\lambda_1
( L_1^2 + L_2^2 + L_3^2 )
( \tilde{\Phi}_1^2 + \tilde{\Phi}_2^2 + \tilde{\Phi}_3^2 )
\nonumber\\
&&
+
\lambda_2
( L_1^2 + \omega^2 L_2^2 + \omega L_3^2 )
( \tilde{\Phi}_1^2 + \omega \tilde{\Phi}_2^2 + \omega^2 \tilde{\Phi}_3^2)
\nonumber\\
&&
+
\lambda_3
( L_1^2 + \omega L_2^2 + \omega^2 L_3^2 )
( \tilde{\Phi}_1^2 + \omega^2 \tilde{\Phi}_2^2 + \omega \tilde{\Phi}_3^2)
\nonumber\\
&&
+
\lambda_4
(
L_2 L_3 \tilde{\Phi}_2 \tilde{\Phi}_3
+
L_3 L_1 \tilde{\Phi}_3 \tilde{\Phi}_1
+
L_1 L_2 \tilde{\Phi}_1 \tilde{\Phi}_2
),
\end{eqnarray}
where $\lambda_1 \ldots \lambda_4$ are complex parameters.

After spontaneous symmetry breaking, the fields get VEVs $\langle \Phi_k
\rangle = v_k$, and the elements of the symmetric matrix $K$ become
\begin{eqnarray}
k_{11} &=&
\lambda_1
( v_1^{\ast 2} + v_2^{\ast 2} + v_3^{\ast 2})
+
\lambda_2
( v_1^{\ast 2} + \omega^2 v_2^{\ast 2} + \omega v_3^{\ast 2})
\nonumber\\
&&
\lambda_3
( v_1^{\ast 2} + \omega v_2^{\ast 2} + \omega^2 v_3^{\ast 2}),
\nonumber\\
k_{22} &=&
\lambda_1
( v_1^{\ast 2} + v_2^{\ast 2} + v_3^{\ast 2})
+
\lambda_2 \omega
( v_1^{\ast 2} + \omega^2 v_2^{\ast 2} + \omega v_3^{\ast 2})
\nonumber\\
&&
\lambda_3 \omega^2
( v_1^{\ast 2} + \omega v_2^{\ast 2} + \omega^2 v_3^{\ast 2}),
\nonumber\\
k_{33} &=&
\lambda_1
( v_1^{\ast 2} + v_2^{\ast 2} + v_3^{\ast 2})
+
\lambda_2 \omega^2
( v_1^{\ast 2} + \omega^2 v_2^{\ast 2} + \omega v_3^{\ast 2})
\nonumber\\
&&
\lambda_3 \omega
( v_1^{\ast 2} + \omega v_2^{\ast 2} + \omega^2 v_3^{\ast 2})
\nonumber\\
k_{12} &=&
\frac{1}{2} v_1^\ast v_2^\ast \lambda_4, \quad
%\nonumber\\
k_{13} =
\frac{1}{2} v_1^\ast v_3^\ast \lambda_4, \quad
%\nonumber\\
k_{23} =
\frac{1}{2} v_2^\ast v_3^\ast \lambda_4.
\end{eqnarray}
Only the VEV alignments $(1,1,1)$ and $(\pm1, \eta, \eta^\ast)$ lead to
nonvanishing charged lepton masses. With $(1,1,1)$, we find
\begin{equation}
K =
\frac{1}{2}
\left(
\begin{array}{ccc}
6 \lambda_1 & \lambda_4 & \lambda_4\\
\lambda_4 & 6 \lambda_1 & \lambda_4\\
\lambda_4 & \lambda_4 & 6 \lambda_1
\end{array}
\right),
\end{equation}
meaning that $K K^\dagger$ has a doubly degenerate eigenvalue $9
|\lambda_1|^2 + |\lambda_4/2|^2 - 3 \textrm{Re}(\lambda_1 \lambda_4^\ast)$
and a third eigenvalue $9 |\lambda_1|^2 + |\lambda_4|^2 + 6
\textrm{Re}(\lambda_1 \lambda_4^\ast)$. With $(\pm1, \eta, \eta^\ast)$ the
matrix $K$ is slightly different, but the eigenvalues of $K K^\dagger$ are
the same with $\lambda_1 \rightarrow \lambda_3$. So, the effective
dimension-5 term with $A_4$ symmetry, $\Phi \sim \mathbf{3}$, and $L_L \sim
\mathbf{3}$ is ruled out.

Finally, we turn to the possibility that $L_L$ is a singlet of $A_4$. We know
from Table~\ref{tabA4:charged} that the only choice compatible with the
charged lepton masses is $L_L \sim (\mathbf{1}, \mathbf{1^\prime},
\mathbf{1^{\prime \prime}})$, $\ell_R \sim \mathbf{3}$. The $L_L L_L$ group
structures obtainable when $L_L \sim (\mathbf{1}, \mathbf{1^\prime},
\mathbf{1^{\prime \prime}})$ are contained in the right-hand side of
Eq.~\eqref{LL_11pp1pp_11p1pp}. Entries with $\mathbf{1}$ couple to the
$\tilde{\Phi} \tilde{\Phi}$ combination $(\mathbf{3} \otimes
\mathbf{3})_\mathbf{1}$. This can be read from Eq.~\eqref{33_1} by changing
$L_L \rightarrow \tilde{\Phi}$, which after spontaneous symmetry breaking
leads to $( v_1^{\ast 2} +  v_2^{\ast 2} + v_3^{\ast 2})$. Similarly, entries
with $\mathbf{1^\prime}$ in Eq.~\eqref{LL_11pp1pp_11p1pp} couple to the
$(\mathbf{3} \otimes \mathbf{3})_\mathbf{1^{\prime \prime}}$ combination $(
v_1^{\ast 2} +  \omega v_2^{\ast 2} + \omega^2 v_3^{\ast 2})$. Finally,
entries with $\mathbf{1^{\prime \prime}}$ in Eq.~\eqref{LL_11pp1pp_11p1pp}
couple to the $(\mathbf{3} \otimes \mathbf{3})_\mathbf{1^\prime}$ combination
$( v_1^{\ast 2} +  \omega^2 v_2^{\ast 2} + \omega v_3^{\ast 2})$.

For the VEV $(1,1,1)$, the above combinations give $v_1^{\ast 2} +  v_2^{\ast
2} + v_3^{\ast 2}=3$, $v_1^{\ast 2} +  \omega v_2^{\ast 2} + \omega^2
v_3^{\ast 2}=0$, and $v_1^{\ast 2} +  \omega^2 v_2^{\ast 2} + \omega v_3^{\ast
2}=0$. In this case, the matrix $K$ is of the form
\begin{equation}
K =
3\,
\left(
\begin{array}{ccc}
\lambda_1 & 0 & 0\\
0 & 0 & \lambda_2\\
0 & \lambda_2 & 0
\end{array}
\right).
\end{equation}
The matrix $K K^\dagger$ has a doubly degenerate eigenvalue $3|\lambda_2|^2$,
and a third eigenvalue $3|\lambda_1|^2$. Similarly, for the VEV $(\pm 1, \eta,
\eta^\ast)$, the relevant VEV combinations are $0$, $3$, and $0$,
respectively, with
\begin{equation}
K =
3 \,
\left(
\begin{array}{ccc}
0 & \lambda_2 & 0\\
\lambda_2 & 0 & 0\\
0 & 0 & \lambda_1
\end{array}
\right).
\end{equation}
Again, $K K^\dagger$ has a doubly degenerate eigenvalue $3|\lambda_2|^2$ and
a third eigenvalue $3|\lambda_1|^2$. As a result, all cases are ruled out.

\subsection{Type-I and type-III seesaw}
\label{sec:IandIII}

We consider first the type-I~\cite{typeI} seesaw mechanism with $n_R=3$
right-handed neutrino fields. The relevant Lagrangian is
\begin{eqnarray}
- {\cal L}_\text{I}
&=&
\bar{L}_L \sum_{k=1}^3 Y_{\ell,k} \Phi_k \ell_R
+ \bar{L}_L \sum_{k=1}^3 Y_{\nu,k}^\ast \tilde{\Phi}_k \nu_R
\nonumber\\
& &
+ \frac{1}{2} \bar{\nu}_R M_R \nu_R^c + \textrm{H.c.},
\label{LY_I}
\end{eqnarray}
where $L_L = (\ell_L, \nu_L)^T$, $\ell_R$, and $\nu_R$ are vectors in the
three-dimensional generation spaces of left-handed doublets, right-handed
charged lepton singlets, and right-handed neutrino singlets, respectively.
For each scalar doublet $\Phi_k$ there is a charged lepton Yukawa matrix
$Y_{\ell, k}$, and a neutrino Yukawa matrix $Y_{\nu, k}$. After the
spontaneous symmetry breaking we obtain the mass terms
\begin{equation}
- {\cal L}_\text{I} =
\bar{\ell}_L m_\ell \ell_R
+ \bar{\nu}_L m_D \nu_R
+ \frac{1}{2} \bar{\nu}_R M_R \nu_R^c + \textrm{H.c.},
\label{L_I}
\end{equation}
where
\begin{equation}
m_\ell = \sum_{k=1}^3 Y_{\ell,k} v_k,
\quad
m_D = \sum_{k=1}^3 Y_{\nu,k}^\ast v_k^\ast,
\end{equation}
and $M_R$ is a symmetric matrix. To correctly reproduce the light neutrino
masses, the eigenvalues of $M_R$ should be much larger than $(|v_1|^2 +
|v_2|^2 + |v_3|^2)^{1/2}$. Integrating out the heavy right-handed Majorana
fields, the low-energy effective Lagrangian becomes
\begin{equation}
- {\cal L}_\text{eff} =
\bar{\ell}_L m_\ell \ell_R
+ \frac{1}{2} \nu_L^T C m_\nu \nu_L
+ \textrm{H.c.},
\label{L_mnu}
\end{equation}
and the light neutrinos acquire an effective mass
\begin{equation}
m_\nu = - m_D M_R^{-1} m_D^T.
\label{m_nu}
\end{equation}
In the basis where the charged lepton mass matrix is diagonal,
\begin{equation}
m_\ell = \textrm{diag}\, (m_e, m_\mu, m_\tau),
\label{diag_mell}
\end{equation}
the neutrino mass matrix $m_\nu$ is diagonalized by the
Pontecorvo-Maki-Nakagawa-Sakata (PMNS)~\cite{PMNS} leptonic mixing matrix $U$ as
\begin{equation}
U^T m_\nu\, U = \textrm{diag}\, (m_{1}, m_{2}, m_{3}),
\label{diagmnu}
\end{equation}
where $m_{i}$ are the light neutrino masses.

We will try to assign the lepton fields to $A_4$ representations, subject to
the following constraints:
\begin{enumerate}
\item The matrix $M_R$ corresponding to the heavy Majorana fields cannot have
a zero eigenvalue;
\item The charged lepton masses cannot vanish or be degenerate.
\end{enumerate}
The first condition forces the right-handed neutrino fields to be in one of the
following three representations. One can have $\nu_R \sim (\mathbf{1},
\mathbf{1}, \mathbf{1})$ and
\begin{equation}
M_R =
\left(
\begin{array}{ccc}
\times & \times & \times\\
\times & \times & \times\\
\times & \times & \times
\end{array}
\right),
\label{MR_111}
\end{equation}
where $\times$ represents an independent complex entry. Alternatively, one can
have $\nu_R \sim (\mathbf{1}, \mathbf{1^{\prime}}, \mathbf{1^{\prime
\prime}})$ and
\begin{equation}
M_R =
\left(
\begin{array}{ccc}
\times & 0 & 0\\
0 & 0 & \times\\
0 & \times & 0
\end{array}
\right).
\label{MR_11p1pp}
\end{equation}
Finally, if $\nu_R \sim \mathbf{3}$, then
\begin{equation}
M_R =
M
\left(
\begin{array}{ccc}
1 & 0 & 0\\
0 & 1 & 0\\
0 & 0 & 1
\end{array}
\right),
\label{MR_3}
\end{equation}
where $M$ is an arbitrary complex number. This corresponds to degenerate
heavy neutrinos. Other combinations are ruled out by our first requirement.

Before proceeding, let us look back at Eq.~\eqref{m_nu}. Because
$\textrm{det}\, M_R \neq 0$, the existence (or absence) of massless light
neutrinos depends on the nature of $m_D$. If $\textrm{det}\,m_D =0$
($\textrm{det}\,m_D \neq 0$), then $\textrm{det}\,m_\nu =0$
($\textrm{det}\,m_\nu\neq 0$). As a result, we will consider the constraints
coming from charged leptons in each of these two cases, separately. Notice
that, regardless of the $\nu_R$ representation, no case with both $L_L$ and
$\ell_R$ in singlet representations is possible because $\Phi \sim
\mathbf{3}$, leading to $m_\ell =0$. Similarly, the cases where both $L_L$
and $\nu_R$ are in singlet representations are excluded because they lead to
$m_D = 0$ and, through Eq.~\eqref{m_nu}, to $m_\nu=0$. Finally, as shown in
Sec.~\ref{sec:A4lepton}, cases when both $L_L$ and $\ell_R$ are in the
triplet representation of $A_4$ are not possible. The remaining cases are
listed in Table~\ref{tabA4:rep}.
\begin{table}[h!]
\begin{center}
\begin{tabular}{ccccccccccc}
\hline\hline
& Case & & $L_L$ & & $\nu_R$ & & $\ell_R$ & & Neutrino masses &\\
\hline\hline
& i) & & $\mathbf{3}$  & &
$\mathbf{3}$  & &
$(\mathbf{1}, \mathbf{1^{\prime}}, \mathbf{1^{\prime \prime}})$ & & 2 degenerate  &\\
& ii) & & $(\mathbf{1}, \mathbf{1^{\prime}}, \mathbf{1^{\prime \prime}})$  & &
$\mathbf{3}$  & &
$\mathbf{3}$ & & 2 degenerate  &\\
%\hline
& iii) & & $\mathbf{3}$  & &
$(\mathbf{1}, \mathbf{1^{\prime}}, \mathbf{1^{\prime \prime}})$  & &
$(\mathbf{1}, \mathbf{1^{\prime}}, \mathbf{1^{\prime \prime}})$ & & 2 degenerate  &\\
& iv) & & $\mathbf{3}$  & &
$(\mathbf{1}, \mathbf{1}, \mathbf{1})$  & &
$(\mathbf{1}, \mathbf{1^{\prime}}, \mathbf{1^{\prime \prime}})$ & & 2 massless  &\\
\hline\hline
\end{tabular}
\end{center}
\caption{\label{tabA4:rep}Possible representations of the left-handed
lepton doublet ($L_L$), the right-handed neutrino singlets ($\nu_R$), and
right-handed charged lepton singlets ($\ell_R$), when the three Higgs
doublets are in a triplet representation $\mathbf{3}$.}
\end{table}

\subsubsection{Nonvanishing neutrino masses}

We start by looking at the cases in which $\nu_R \sim \mathbf{3}$. From
Table~\ref{tabA4:rep} one concludes that there are two possibilities which
may lead to nonvanishing neutrino and charged lepton masses: i) $L_L \sim
\mathbf{3}$, $\nu_R \sim \mathbf{3}$, $\ell_R \sim (\mathbf{1},
\mathbf{1^{\prime}}, \mathbf{1^{\prime \prime}})$ and ii) $L_L \sim
(\mathbf{1}, \mathbf{1^{\prime}}, \mathbf{1^{\prime \prime}})$, $\nu_R \sim
\mathbf{3}$, $\ell_R \sim \mathbf{3}$.

Let us consider in detail case i). Then, Eq.~\eqref{MR_3} holds and we may
parametrize
\begin{equation}
m_\ell =
\left(
\begin{array}{ccc}
a\, e^{i \alpha}\, v_1 & b\, e^{i \beta}\, v_1 & c\, e^{i \gamma}\, v_1\\
a\, e^{i \alpha}\, v_2 & \omega\, b\, e^{i \beta}\, v_2 & \omega^2\, c\, e^{i \gamma}\, v_2\\
a\, e^{i \alpha}\, v_3 & \omega^2\, b\, e^{i \beta}\, v_3 & \omega\, c\, e^{i \gamma}\, v_3
\end{array}
\right),
\label{ml_331}
\end{equation}
and
\begin{equation}
m_D =
\left(
\begin{array}{ccc}
0 & f\, e^{i \epsilon}\, v_3^\ast & d\, e^{i \delta}\, v_2^\ast\\
d\, e^{i \delta}\, v_3^\ast & 0 & f\, e^{i \epsilon}\, v_1^\ast\\
f\, e^{i \epsilon}\, v_2^\ast & d\, e^{i \delta}\, v_1^\ast  & 0
\end{array}
\right).
\label{mD_331}
\end{equation}
Taking the VEV alignment $(1,1,1)$ or $(\pm 1, \eta, \eta^*)$, the
eigenvalues of $m_\ell m_\ell^\dagger$ are $3 a^2$, $3 b^2$, and $3 c^2$,
which can be properly chosen to fit the experimental values of the
charged lepton masses.

As for the light neutrino mass matrix, we obtain for the VEV alignment
$(1,1,1)$
\begin{equation}
m_\nu = -M^{-1} m_D m_D^T
= -M^{-1}\left(
\begin{array}{ccc}
x
& y
& y\\
y
& x
& y\\
y
& y
& x
\end{array}
\right),
\end{equation}
where $x=d^2\, e^{2 i \delta} + f^2\, e^{2 i \epsilon}$ and $y=d\, f\, e^{i
(\delta + \epsilon)}$. The eigenvalues of $m_\nu m_\nu^\dagger$ are
\begin{eqnarray}
&&
M^{-2} \left[d^2 + f^2 + 2 d f \cos{(\delta - \epsilon)}\right]^2,
\\
&&
M^{-2}\left\{\left[d^2 + f^2 - d f \cos{(\delta - \epsilon)}\right]^2
- 3 d^2 f^2 \sin^2(\delta - \epsilon)\right\},\nonumber
\end{eqnarray}
with the latter twice degenerate. This in turn implies that two light
neutrinos are degenerate in mass, in contradiction with experiment. This
feature remains for the VEV alignment $(\pm 1, \eta, \eta^*)$, although the
expressions for the eigenvalues become more involved in that case.

Let us now analyze case ii). In this case, the mass matrices become
\begin{eqnarray}
m_\ell &=&
\left(
\begin{array}{ccc}
a\, e^{i \alpha}\, v_1 & a\, e^{i \alpha}\, v_2 & a\, e^{i \alpha}\, v_3\\
b\, e^{i \beta}\, v_1 & \omega\, b\, e^{i \beta}\, v_2 & \omega^2\, b\, e^{i \beta}\, v_3\\
c\, e^{i \gamma}\, v_1 & \omega^2\, c\, e^{i \gamma}\, v_2 & \omega\, c\, e^{i \gamma}\, v_3
\end{array}
\right),
\label{mell_Ls_ell3}
\\
m_D &=&
\left(
\begin{array}{ccc}
d\, e^{i \delta}\, v_1^\ast & d\, e^{i \delta}\, v_2^\ast
& d\, e^{i \delta}\, v_3^\ast\\
f\, e^{i \epsilon}\, v_1^\ast & \omega\, f\, e^{i \epsilon}\, v_2^\ast
& \omega^2\, f\, e^{i \epsilon}\, v_3^\ast\\
g\, e^{i \xi}\, v_1^\ast & \omega^2\, g\, e^{i \xi}\, v_2^\ast
& \omega\, g\, e^{i \xi}\, v_3^\ast
\end{array}
\right).
\label{mD_Ls_ell3}
\end{eqnarray}
For both VEVs, $(1,1,1)$ and $(\pm 1, \eta, \eta^*)$, $m_\ell m_\ell^\dagger =
3\, \textrm{diag}\, (a^2, b^2, c^2)$, so we can easily accommodate the
charged lepton masses. Furthermore, for the VEV $(1,1,1)$, we obtain
\begin{equation}
m_\nu =  - M^{-1} m_D m_D^T
=
-M^{-1}
\left(
\begin{array}{ccc}
z
& 0
& 0\\
0
& 0
& t\\
0
& t
& 0
\end{array}
\right),
\end{equation}
where $z = 3 d^2\, e^{2 i \delta}$ and $t = 3 f g\, e^{i (\epsilon + \xi)}$.
Thus, $m_\nu m_\nu^\dagger = 9\,\textrm{diag}\, (d^4, f^2 g^2, f^2 g^2)$, and
we get two degenerate light neutrinos. The matrices for the VEV $(\pm 1,
\eta, \eta^*)$ are slightly different, but the conclusions are the same. As a
result, cases i) and ii) are ruled out by experiment.

The analysis of the remaining case iii), for which $\nu_R \sim (\mathbf{1},
\mathbf{1^{\prime}}, \mathbf{1^{\prime \prime}})$, $L_L \sim \mathbf{3}$ and
$\ell_R \sim (\mathbf{1}, \mathbf{1^{\prime}}, \mathbf{1^{\prime \prime}})$,
is easily carried out. Indeed, the charged lepton sector coincides with that
of case i). Furthermore, from Eq.~\eqref{ml_331}, the matrix $m_D$ can be
inferred:
\begin{equation}
m_D =
\left(
\begin{array}{ccc}
d\, e^{i \delta}\, v_1^\ast & f\, e^{i \epsilon}\, v_1^\ast
& g\, e^{i \xi}\, v_1^\ast\\
d\, e^{i \delta}\, v_2^\ast & \omega\, f\, e^{i \epsilon}\, v_2^\ast
& \omega^2\, g\, e^{i \xi}\, v_2^\ast\\
d\, e^{i \delta}\, v_3^\ast & \omega^2\, f\, e^{i \epsilon}\, v_3^\ast
& \omega\, g\, e^{i \xi}\, v_3^\ast
\end{array}
\right).
\label{mD_313}
\end{equation}
Writing
\begin{equation}
M_R =
\left(
\begin{array}{ccc}
r_1\, e^{i \sigma_1} & 0 & 0\\
0 & 0 & r_2\, e^{i \sigma_2}\\
0 & r_2\, e^{i \sigma_2} & 0
\end{array}
\right),
\end{equation}
we can use Eq.~\eqref{m_nu} to determine $m_\nu$. The expression is long, but
the eigenvalues of $m_\nu\, m_\nu^\dagger$ are simply given by $(3
d^2/r_1)^2$ and $(3 f g /r_2)^2$, with the latter twice degenerate. Thus, we
also get two degenerate light neutrinos, so this case is also excluded.

\subsubsection{Vanishing neutrino masses}

We now turn to the possibility that $m_D$, and thus $m_\nu$, have
determinants equal to zero, with at most one massless light neutrino. The
only case consistent with realistic charged lepton masses is iv) $L_L \sim
\mathbf{3}$, $\nu_R \sim (\mathbf{1}, \mathbf{1}, \mathbf{1})$, $\ell_R \sim
(\mathbf{1}, \mathbf{1^{\prime}}, \mathbf{1^{\prime \prime}})$, provided that
the VEV alignment is $(1,1,1)$ or $(\pm 1, \eta, \eta^*)$.

Let us show that this possibility is also inconsistent with experiment. In
this case, $m_\ell$ has the same matrix structure as case i), while the Dirac
neutrino mass matrix has the form
\begin{equation}
m_D
=
\left(
\begin{array}{ccc}
d\, e^{i \delta}\, v_1^\ast & f\, e^{i \epsilon}\, v_1^\ast
& g\, e^{i \xi}\, v_1^\ast\\
d\, e^{i \delta}\, v_2^\ast & f\, e^{i \epsilon}\, v_2^\ast
& g\, e^{i \xi}\, v_2^\ast\\
d\, e^{i \delta}\, v_3^\ast & f\, e^{i \epsilon}\, v_3^\ast
& g\, e^{i \xi}\, v_3^\ast
\end{array}
\right).
\end{equation}
Taking the VEV $(1,1,1)$, we find that
\begin{equation}
m_D = V_L\ D_D\ V_R^\dagger\,,
\end{equation}
where $D_D = \sqrt{3}\, e^{i \xi} R_3\, \textrm{diag} (0,0,1)$,
\begin{equation}
V_L
=
\left(
\begin{array}{ccc}
-\frac{1}{\sqrt{2}} & \frac{1}{\sqrt{6}} & \frac{1}{\sqrt{3}}\\*[1mm]
0 & - \sqrt{\frac{2}{3}} & \frac{1}{\sqrt{3}}\\*[1mm]
\frac{1}{\sqrt{2}} & \frac{1}{\sqrt{6}} & \frac{1}{\sqrt{3}}\\
\end{array}
\right),
\end{equation}
\begin{equation}
V_R
=
\left(
\begin{array}{ccc}
- \frac{g}{R_2}\, e^{-i(\delta-\xi)}
& \frac{d f}{R_2 R_3}\, e^{i(\epsilon-\xi)}
& \frac{d}{R_3}\, e^{-i(\delta-\xi)}
\\*[3mm]
0
& - \frac{R_2}{R_3}\, e^{i(\delta-\xi)}
& \frac{f}{R_3}\, e^{-i(\epsilon-\xi)}
\\*[3mm]
\frac{d}{R_2}
& \frac{f g}{R_2 R_3}\, e^{i(\delta + \epsilon - 2\xi)}
&  \frac{g}{R_3}\\
\end{array}
\right),
\end{equation}
$R_2=(d^2 + g^2)^{1/2}$ and $R_3 = (d^2 + f^2 + g^2)^{1/2}$. Here, $V_L$ and
$V_R$ are the unitary matrices that diagonalize the Hermitian matrices $m_D
m_D^\dagger$ and $m_D^\dagger m_D$, respectively.

Using Eq.~\eqref{m_nu}, we get
\begin{equation}
m_\nu
=
- V_L\, D_D\, X\, D_D\, V_L^T
=
- e^{2 i \xi} R_3^2 X_{33}
\left(
\begin{array}{ccc}
1 & 1 & 1\\
1 & 1 & 1\\
1 & 1 & 1
\end{array}
\right),
\label{mnu_313_v1}
\end{equation}
where $X = V_R^\dagger\, M_R^{-1}\, V_R^\ast$.

From Eq.~\eqref{mnu_313_v1}, we find the eigenvalues of $m_\nu m_\nu^\dagger$
to be $m^2_{1} = m^2_{2} = 0$ and $m^2_{3} = 9 R_3^4\, |X_{33}|^2$. Since there
are two massless neutrinos, this case is ruled out. For the VEV alignment $(1,
\eta, \eta^\ast)$, the intermediate steps get more involved, but the
eigenvalues of $m_\nu m_\nu^\dagger$ have the same expressions, and therefore
this possibility is also ruled out.

Before concluding this section, let us comment on the type-III seesaw
mechanism~\cite{typeIII}. In the type-III seesaw framework, instead of three
right-handed singlet neutrino fields, one adds three Majorana neutrinos,
$\Sigma_R$, in the triplet representation of the gauge group $SU(2)_L$,
\begin{equation}
\Sigma_{iR}
=
\left(
\begin{array}{cc}
\Sigma_i^0/\sqrt{2} & \Sigma_i^+ \\
\Sigma_i^- & - \Sigma_i^0/\sqrt{2}
\end{array}
\right),\quad i = 1, 2, 3.
\end{equation}
The relevant Lagrangian is very similar to Eq.~\eqref{LY_I} for a type-I seesaw:
\begin{eqnarray}
- {\cal L}_\text{II}
&=&
\bar{L}_L \sum_{k=1}^3 Y_{\ell,k} \Phi_k \ell_R
+ \bar{L}_L \sum_{k=1}^3 Y_{\Sigma,k}^\ast \tilde{\Phi}_k \Sigma_R
\nonumber\\
& &
+ \frac{1}{2} \sum_{i,j=1}^3 (M_\Sigma)_{ij}
\textrm{Tr}\left(\bar{\Sigma}_{iR}  \Sigma_{jR}^C\right) + \textrm{H.c.}
\label{LY_III}
\end{eqnarray}
The effective light neutrino mass matrix acquires the same seesaw structure as
Eq.~\eqref{m_nu}, with $M_R$ replaced by $M_\Sigma$. As a result, the analysis
of flavor structures under the $A_4$ symmetry is the same as before, and all
the conclusions hold. In particular, Table~\ref{tabA4:rep} applies, with the
obvious replacement $\nu_R \rightarrow \Sigma_R$.

\section{Softly broken $A_4$ symmetry}
\label{sec:soft}

We now consider the possibility that the effective operator is not invariant
under $A_4$. This situation is well behaved as long as we guarantee that the
non-invariance comes, at the UV level, from terms that do not spoil
renormalizability. We therefore assume that $A_4$ is broken
softly by dimension-3 terms contributing to the right-handed neutrino
mass matrix $M_R$~\cite{Branco:2009by}.

We start again from the cases listed in Table~\ref{tabA4:rep} and analyzed
in Sec.~\ref{sec:IandIII}. Let us consider first case iv) of
Table~\ref{tabA4:rep}, where $\nu_R \sim (\mathbf{1}, \mathbf{1},
\mathbf{1})$, and $M_R$ is, according to Eq.~\eqref{MR_111}, the most general
$3 \times 3$ symmetric complex matrix. Including soft-breaking terms does not
alter this feature. Since Eq.~\eqref{mnu_313_v1} leads to two massless
eigenvalues in $m_\nu m_\nu^\dagger$, we conclude that case iv) is not
viable, even after the inclusion of soft-breaking terms.

Next, we show that the remaining cases listed in Table~\ref{tabA4:rep} lead to
viable fits of the current experimental neutrino data after the inclusion of
soft-breaking terms in $M_R$.

We start with case i). The charged lepton and Dirac neutrino mass matrices
are given by Eqs.~\eqref{ml_331} and~\eqref{mD_331}, respectively. For
simplicity, we study the VEV alignment $(1,1,1)\,$---the results for $(\pm 1,
\eta, \eta^*)$ will be equivalent. We can always change the basis of $L_L$,
corresponding to multiplying the matrices $m_\ell$ and $m_D$ on the left by
the same unitary matrix. In this case, the matrices $m_\ell m_\ell^\dagger$
and $m_D m_D^\dagger$ are diagonalized by the same matrix,
\begin{equation}
V_\omega
=
\frac{1}{\sqrt{3}}
\left(
\begin{array}{ccc}
1 & 1 & 1\\
1 & \omega & \omega^2\\
1 & \omega^2 & \omega
\end{array}
\right).
\label{V_omega}
\end{equation}
Indeed, multiplying the mass matrices on the left by $V_\omega^\dagger$, we
obtain in the new basis
\begin{eqnarray}
m_\ell
&=&
\sqrt{3}\
\textrm{diag}
\left(
a\, e^{i\, \alpha},
b\, e^{i\, \beta},
c\, e^{i\, \gamma}
\right),
\nonumber\\
m_D
&=&
D_D\, V_R^\dagger\,,
\end{eqnarray}
where
\begin{eqnarray}
V_R^\dagger
&=&
\frac{1}{\sqrt{3}}
\left(
\begin{array}{ccc}
1 & 1 & 1\\
\omega^2 & \omega & 1\\
\omega & \omega^2 & 1
\end{array}
\right),
\\
D_D
&=& \text{diag}\,(d e^{i\delta} + f e^{i\epsilon}, d e^{i\delta} + \omega^2 f e^{i\epsilon},
d e^{i\delta} + \omega f e^{i\epsilon}).\nonumber
\end{eqnarray}
The PMNS mixing matrix and the light neutrino masses are then obtained from the
diagonalization of $m_\nu$ [cf. Eqs.~\eqref{m_nu} and \eqref{diagmnu}].

We have randomly generated matrices $m_D$ and $M_R$ satisfying the current
experimental data, via a procedure described in detail in Appendix~\ref{fit}.
In particular, we define in Eq.~\eqref{sigma1} a figure of merit $\sigma$
probing how much the matrix $M_R$ differs from its form in the exact $A_4$
limit, given in the present case by Eq.~\eqref{MR_3}. Values of $\sigma \sim
1$ correspond to soft-breaking terms of the order of the $A_4$-symmetric
terms. Smaller values of $\sigma$ correspond to cases where the terms
that break the symmetry are perturbative; that is, they are about 1 order
of magnitude smaller than the terms which preserve $A_4$.
\begin{figure}[htb]
\includegraphics*[height=6cm]{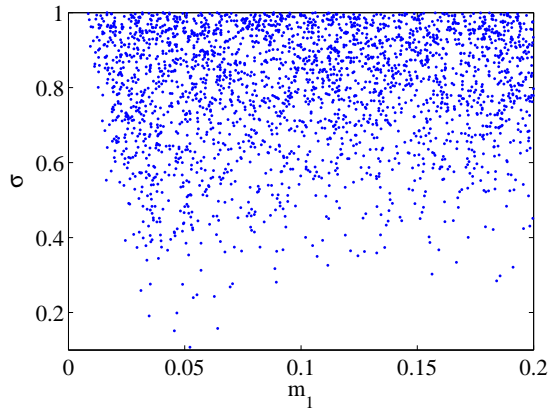}
\caption{\label{figure1}Simulation points in case i), showing how the figure
of merit $\sigma$ varies with $m_1$ (in eV). The quantity $\sigma$ measures
the deviation of $M_R$, in the softly broken case, from its form in the
$A_4$-symmetric case.}
\end{figure}

Our aim is to show that there exist viable solutions where the deviations from
Eq.~\eqref{MR_3} are perturbative. This becomes clear from Fig.~\ref{figure1},
which shows $\sigma$ as a function of $m_1$ (in eV). We notice several
features: a) one can produce fits with values of $m_1$ in almost the whole
range attempted (from 0 to 0.2~eV), except for very small values of $m_1$;
b) solutions with $\sigma < 0.2$ can be found, meaning that the soft-breaking
terms are perturbative; c) smaller values of $\sigma$ tend to prefer values
for $m_1$ around $0.05$~eV.

\begin{figure}[htb]
\includegraphics*[height=6cm]{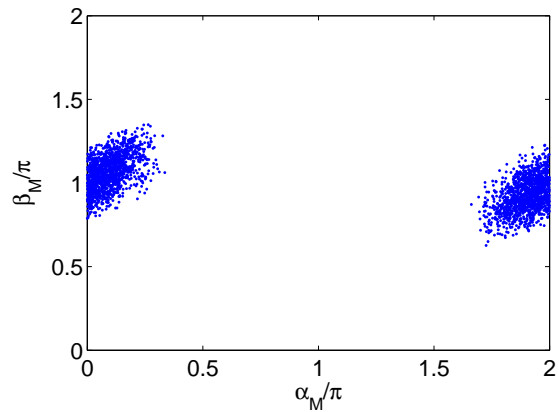}
\caption{\label{figure2}Simulation in the $(\alpha_M,\beta_M)$ plane for case
i). Points verifying the condition $\sigma < 1$ are represented. }
\end{figure}

In our simulation, smaller values of $\sigma$ have no correlation with the
measured quantities---namely, the neutrino mixing angles $\theta_{12}$,
$\theta_{13}$, $\theta_{23}$, and the mass-squared differences $\Delta
m^2_{21} =m_2^2 - m_1^2$, or $\Delta m^2_{31} =m_3^2 - m_1^2$. Although
$\sigma < 0.2$ implies values of the Dirac phase $\delta_D$ close to $\pm
\pi/2$, all values of $\delta_D$ are possible if one allows $\sigma < 1$. In
contrast, even a loose cut of $\sigma < 1$, forces the Majorana phases
$\alpha_M \sim 0$ and $\beta_M \sim \pi$ (recall that the phases are defined
mod $2\pi$). This is illustrated in Fig.~\ref{figure2}.

Next, we consider case ii), where $L_L \sim (\mathbf{1}, \mathbf{1^{\prime}},
\mathbf{1^{\prime \prime}})$, $\nu_R \sim \mathbf{3}$, $\ell_R \sim
\mathbf{3}$, so that Eqs.~\eqref{mell_Ls_ell3} and \eqref{mD_Ls_ell3} hold for
the charged lepton mass matrix $m_\ell$ and the Dirac neutrino mass matrix
$m_D$, respectively. We choose the VEV alignment $(1, 1, 1)$. This case is
interesting because $m_\ell$ is diagonalized exclusively through a unitary
transformation on the right-handed fields $\ell_R$, implying that the PMNS
matrix arises exclusively from the diagonalization of $m_\nu$. The eigenvalues
of $m_\ell\, m_\ell^\dagger$ from Eq.~\eqref{mell_Ls_ell3} are $3a^2$, $3b^2$,
and $3c^2$. Thus, we take $a= m_e/\sqrt{3}$, $b= m_\mu/\sqrt{3}$, and $c=
m_\tau/\sqrt{3}$, and consider $\alpha = \beta = \gamma = 0$.

We will now assume that $A_4$ is softly broken in the right-handed neutrino
sector, such that the form of $M_R$ in Eq.~\eqref{MR_3} is altered. As before,
we follow the fit procedure described in Appendix~\ref{fit}.
Figure~\ref{figure3} shows $\sigma$ as a function of $s_{23}^2 =
\sin^2{\theta_{23}}$.
\begin{figure}[htb]
\includegraphics*[height=6cm]{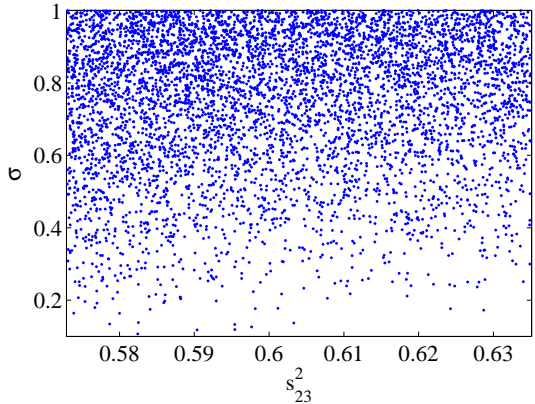}
\caption{\label{figure3}Simulation points for case ii), showing how $\sigma$
varies with $s_{23}^2$. }
\end{figure}
We notice that all values of $s_{23}^2$ are possible, but that smaller values
of $\sigma$ show some preference for smaller values of $s_{23}^2$. There is no
such correlation with $\theta_{12}$, $\theta_{13}$, $m_1$, $\Delta m^2_{21}$,
or $\Delta m^2_{31}$. The CP-violating phases $\delta_D$, $\alpha_M$, and
$\beta_M$ exhibit the same behavior as in case i).

It is interesting to look at a numerical example in detail. We consider the
symmetric matrix $M_R = M Y_R$, with $M=10^{14}$~GeV and
\begin{eqnarray}
(Y_R)_{11} &=&0.6622 + 7.7769\, i,\nonumber\\
(Y_R)_{12} &=&-0.4304 + 0.5404\, i,\nonumber\\
(Y_R)_{13} &=&-0.0490 - 0.4532\, i,\\
(Y_R)_{22} &=&5.0726 + 5.9385\, i,\nonumber\\
(Y_R)_{23} &=&0.1819 + 0.1479\, i,\nonumber\\
(Y_R)_{33} &=&6.5927 + 4.2281\, i.\nonumber
\end{eqnarray}
For use in the matrix $m_D$ of Eq.~\eqref{mD_Ls_ell3}, we take
\begin{eqnarray}
d\, e^{i \delta} &=&0.7663 + 0.2958\, i,
\nonumber\\
f\, e^{i \epsilon} &=&0.7516 - 0.0537\, i,
\nonumber\\
g\, e^{i \xi} &=&-0.7477 - 0.5013\, i,
\end{eqnarray}
and $v_1=v_2=v_3=246$~GeV. Upon diagonalization of the neutrino mass matrix
$m_\nu$ given by Eq.~\eqref{m_nu}, we find a nearly degenerate normal neutrino
mass spectrum with $m_{1}\simeq 0.1545$~eV, $\Delta m^2_{21} = 7.46 \times
10^{-5}\, \textrm{eV}^2$ and $\Delta m^2_{31} = 2.55 \times 10^{-3}\,
\textrm{eV}^2$, which are well within the $1\sigma$ ranges given in
Ref.~\cite{Tortola:2012te}. Next, we equate the diagonalizing matrix $U$ [see
Eq.~\eqref{diagmnu}] with the PMNS neutrino mixing matrix written in the
standard PDG form~\cite{pdg}. This leads to $\sin^2{\theta_{12}} = 0.336$,
$\sin^2{\theta_{23}} = 0.613$, and $\sin^2{\theta_{13}} = 0.0247$. According
to Ref.~\cite{Tortola:2012te}, the first value is close to its $1 \sigma$
upper bound, while the other two coincide with the best-fit values. Finally,
the as-yet-unmeasured CP-violating phases turn out to be $\delta = 0.53 \pi$,
$\alpha_M = 0.28 \pi$, and $\beta_M = 1.08 \pi$.

In the above numerical example, the magnitudes of the entries in $M_R$ are
\begin{equation}
|M_R| =
7.83 \times 10^{14}\, \textrm{GeV}\times
\left(
\begin{array}{ccc}
0.997 & 0.088 & 0.058\\
0.088 & 0.997 & 0.030 \\
0.058 & 0.030 & 1.000
\end{array}
\right).
\end{equation}
This shows that the current neutrino oscillation data can be easily fitted in
flavor models of the type discussed in this article, if the $A_4$ symmetry is
softly broken in the heavy Majorana neutrino mass matrix $M_R$ by coefficients
that deviate perturbatively (of the order of $10\%$ or less) from the leading
$A_4$-symmetric terms. Notice also that in the above example a normal
(quasidegenerate) neutrino mass spectrum is obtained. Numerical examples that
lead to an inverted neutrino mass hierarchy, and simultaneously to low-energy
neutrino parameters in agreement with the present $1 \sigma$ ranges, can be
equally constructed.

Finally, we turn to case iii), where $L_L \sim \mathbf{3}$, $\nu_R \sim
(\mathbf{1}, \mathbf{1^{\prime}}, \mathbf{1^{\prime \prime}})$, $\ell_R \sim
(\mathbf{1}, \mathbf{1^{\prime}}, \mathbf{1^{\prime \prime}})$. The
charged lepton mass matrix is given in Eq.~\eqref{ml_331}, while the Dirac
neutrino mass matrix $m_D$ is given by Eq.~\eqref{mD_313}. As before, we
concentrate on the VEV $(1,1,1)$. Using Eq.~\eqref{V_omega}, we multiply the
matrices $m_\ell$ and $m_D$ by $V_\omega^\dagger$, obtaining, in the new
basis,
\begin{eqnarray}
m_\ell
&=&
\sqrt{3}\
\textrm{diag}
\left(
a\, e^{i\, \alpha},
b\, e^{i\, \beta},
c\, e^{i\, \gamma}
\right),
\nonumber\\
m_D
&=&
\sqrt{3}\
\textrm{diag}
\left(
d\, e^{i\, \delta},
f\, e^{i\, \epsilon},
g\, e^{i\, \xi}
\right).
\end{eqnarray}
The numerical results for this case exhibit many similarities with case ii).
In particular, there are no fundamental constraints on the observables, except
for $\delta_D \sim \pm \pi/2$, $\alpha_M \sim 0$, and $\beta_M \sim \pi$.

\section{Conclusions}
\label{sec:conclusions}

In this paper we have studied the possibility of generating the neutrino
masses and mixing in the context of models with three scalar doublets in the
triplet representation of the $A_4$ group and three lepton families. We have
shown that none of the possible VEV alignments that correspond to a global
minimum of the scalar potential yields phenomenologically viable
charged lepton and neutrino mass matrices. In particular, there is no
representation assignment that leads to a dimension-5 neutrino mass
operator consistent with the present oscillation data. This in turn implies
that, in this minimal $A_4$ construction, the canonical type-I (type-III)
seesaw mechanism is not consistent with experiment. Notice that, from the
point of view of the low-energy effective operator, this conclusion holds for
any number of right-handed singlet (triplet) neutrinos, since the
dimension-5 operator is the same regardless of the number of heavy fields.
Furthermore, since $A_4$ is a subgroup of $S_4$, our conclusions also remain
valid in flavor models based on the latter group.

In the context of a type-I seesaw mechanism with three heavy right-handed
neutrinos, we have analyzed in detail what happens when the $A_4$-symmetric
Lagrangian is enlarged by adding soft breaking, through dimension-3
right-handed neutrino mass terms. We find three cases where this framework can
be implemented in good agreement with neutrino oscillation data. We have also
pointed out that this is possible perturbatively, i.e., by keeping
the soft-breaking terms much smaller than the $A_4$-symmetric terms of the
right-handed neutrino mass matrix.

At this point, it is worthwhile to comment on other possibilities considered
in the literature. Several studies propose various renormalizable extensions
of the $A_4$ three-Higgs-doublet model, including always additional fields as
well as new symmetries. For example, Ma and Rajasekaran \cite{Ma:2001dn}
discuss a very simple extension, where an additional scalar doublet $\eta$ is
added to case i). This scalar will be  the one responsible for the Dirac
neutrino mass term. Therefore, an additional $Z_2$ is added in order to forbid
$\Phi$ and $\eta$ to interchange sectors. This model still needs some
soft-breaking terms to split the degeneracy in the light neutrino spectrum.
There are also models using two scalar $A_4$ triplets, one for each sector. In
most models these are flavon fields, which lead to a non-renormalizable theory
\cite{Altarelli:2005yx, Chen:2009um}. Renormalizable models with two scalar
triplets of $A_4$ will lead to a six-Higgs-doublet model, with a large
increase in the number of parameters. In Ref.~\cite{Frampton:2008ci} a case
iii) is studied, but with a Higgs triplet for each sector, and with an
additional $Z_2$ symmetry. However, the vacuum alignments utilized ---
$(v_1,v_2,v_3)$ and $v(1,1,-2)$ --- are not shown to be the absolute minimum.
To our knowledge, our work is the first to fully study all possibilities
consistent with one Higgs $A_4$ triplet, for which the global minima of
the scalar potential have been recently identified.

The analysis performed in this paper dealt with the most general case of
three-Higgs-doublet models with the scalars in a triplet of $A_4$. This study
can serve as a starting point for more elaborate models. Nevertheless, we
emphasize that the scalar potential in such new extensions has to be fully
analyzed, in order to ensure that the vacua utilized can indeed be global
minima. In general, this is not a trivial task.

\begin{acknowledgments}
The work of R.G.F. and J.P.S. was partially supported by Portuguese national
funds through FCT - \textit{Funda\c{c}\~{a}o para a Ci\^{e}ncia e a Tecnologia}, under
Projects PEst-OE/FIS/UI0777/2011, PTDC/FIS/098188/2008 and
CERN/FP/116328/2010, and by the EU RTN Marie Curie Project
PITN-GA-2009-237920. The work of H.S. is funded by the European FEDER and
Spanish MINECO, under Grant No. FPA2011-23596.
\end{acknowledgments}

\appendix
\section{Couplings of $L_L L_L$ when $L_L$ is in a singlet representation of $A_4$}
\label{appA4}
Let us consider the group structure of the combination $L_L L_L$, when $L_L$
is an $A_4$ singlet. We concentrate on the group constraints, ignoring any
spinor or $SU(2)_L$ characteristics. We also note that, in all cases of
interest to us, the $L_L L_L$ coupling matrix \textit{must be symmetric}.
Disregarding irrelevant permutations, we must consider the cases

\begin{eqnarray}
(\mathbf{1}, \mathbf{1}, \mathbf{1}),
& \quad &
(\mathbf{1}, \mathbf{1}, \mathbf{1^\prime}),
\nonumber\\
(\mathbf{1}, \mathbf{1^\prime}, \mathbf{1^\prime}),
& \quad &
(\mathbf{1}, \mathbf{1}, \mathbf{1^{\prime \prime}}),
\nonumber\\
(\mathbf{1}, \mathbf{1^{\prime \prime}}, \mathbf{1^{\prime \prime}}),
& \quad &
(\mathbf{1}, \mathbf{1^\prime}, \mathbf{1^{\prime \prime}}),
\nonumber\\
(\mathbf{1^\prime}, \mathbf{1^\prime}, \mathbf{1^\prime}),
& \quad &
(\mathbf{1^\prime}, \mathbf{1^\prime}, \mathbf{1^{\prime\prime}}),
\nonumber\\
(\mathbf{1^\prime}, \mathbf{1^{\prime\prime}}, \mathbf{1^{\prime\prime}}),
& \quad &
(\mathbf{1^{\prime\prime}}, \mathbf{1^{\prime\prime}}, \mathbf{1^{\prime\prime}}).
\label{s_possibilities}
\end{eqnarray}

We can combine the couplings of $L_L L_L$ with some other group structure in
a $3 \times 3$ matrix. To explain our notation, we will use the example of
$L_L \sim (\mathbf{1}, \mathbf{1}, \mathbf{1})$. We construct the matrix of
all field products
\begin{equation}
\left(
\begin{array}{ccc}
\mathbf{1} & \mathbf{1} & \mathbf{1}\\
\mathbf{1} & \mathbf{1} & \mathbf{1}\\
\mathbf{1} & \mathbf{1} & \mathbf{1}
\end{array}
\right),
\label{LL_111}
\end{equation}
where a matrix element $\mathbf{1}$ means that there is at that matrix
position an arbitrary complex entry, if we are coupling $L_L L_L$ to some
other group structure transforming like $\mathbf{1}$ of $A_4$, and zero
otherwise. For example, the scalar combination $(\Phi\Phi) \sim \mathbf{1}$
has couplings to all bilinears of $L_L L_L$ when $L_L \sim (\mathbf{1},
\mathbf{1}, \mathbf{1})$, while a $(\Phi\Phi) \sim \mathbf{1^\prime}$ will
couple to none.

As a further example, consider $L_L \sim (\mathbf{1}, \mathbf{1},
\mathbf{1^\prime})$. The corresponding matrix is
\begin{equation}
\left(
\begin{array}{ccc}
\mathbf{1} & \mathbf{1} & \mathbf{1^\prime}\\
\mathbf{1} & \mathbf{1} & \mathbf{1^\prime}\\
\mathbf{1^\prime} & \mathbf{1^\prime} & \mathbf{1^{\prime \prime}}
\end{array}
\right).
\label{LL_111p}
\end{equation}
This means that a $(\Phi\Phi) \sim \mathbf{1}$ will only introduce couplings
in the upper-left $2 \times 2$ submatrix, and a $(\Phi\Phi) \sim
\mathbf{1^{\prime \prime}}$ will only have couplings to $(L_i L_3 + L_3 L_i)$
with $i=1,2$, while  $(\Phi\Phi) \sim \mathbf{1^\prime}$ would only couple to
$L_3 L_3$. For simplicity, we write $L_L = (L_1, L_2, L_3)$. We also recall
that $\mathbf{1^\prime} \otimes \mathbf{1^{\prime \prime}} = \mathbf{1}$.

The remaining possibilities in Eq.~\eqref{s_possibilities} lead to
\begin{eqnarray}
\left(
\begin{array}{ccc}
\mathbf{1} & \mathbf{1^\prime} & \mathbf{1^\prime}\\
\mathbf{1^\prime} & \mathbf{1^{\prime \prime}} & \mathbf{1^{\prime \prime}}\\
\mathbf{1^\prime} & \mathbf{1^{\prime \prime}} & \mathbf{1^{\prime \prime}}
\end{array}
\right),
& \quad &
\left(
\begin{array}{ccc}
\mathbf{1} & \mathbf{1} & \mathbf{1^{\prime \prime}}\\
\mathbf{1} & \mathbf{1} & \mathbf{1^{\prime \prime}}\\
\mathbf{1^{\prime \prime}} & \mathbf{1^{\prime \prime}} & \mathbf{1^\prime}
\end{array}
\right),
\label{LL_11p1p_111pp}
\\
\left(
\begin{array}{ccc}
\mathbf{1} & \mathbf{1^{\prime \prime}} & \mathbf{1^{\prime \prime}}\\
\mathbf{1^{\prime \prime}} & \mathbf{1^\prime} & \mathbf{1^\prime}\\
\mathbf{1^{\prime \prime}} & \mathbf{1^\prime} & \mathbf{1^\prime}
\end{array}
\right),
& \quad &
\left(
\begin{array}{ccc}
\mathbf{1} & \mathbf{1^\prime} & \mathbf{1^{\prime \prime}}\\
\mathbf{1^\prime} & \mathbf{1^{\prime \prime}} & \mathbf{1}\\
\mathbf{1^{\prime \prime}} & \mathbf{1} & \mathbf{1^\prime}
\end{array}
\right),
\label{LL_11pp1pp_11p1pp}
\\
\left(
\begin{array}{ccc}
\mathbf{1^{\prime \prime}} & \mathbf{1^{\prime \prime}} & \mathbf{1^{\prime \prime}}\\
\mathbf{1^{\prime \prime}} & \mathbf{1^{\prime \prime}} & \mathbf{1^{\prime \prime}}\\
\mathbf{1^{\prime \prime}} & \mathbf{1^{\prime \prime}} & \mathbf{1^{\prime \prime}}
\end{array}
\right),
& \quad &
\left(
\begin{array}{ccc}
\mathbf{1^{\prime \prime}} & \mathbf{1^{\prime \prime}} & \mathbf{1}\\
\mathbf{1^{\prime \prime}} & \mathbf{1^{\prime \prime}} & \mathbf{1}\\
\mathbf{1} & \mathbf{1} & \mathbf{1^\prime}
\end{array}
\right),
\label{LL_1p1p1p_1p1p1pp}
\\
\left(
\begin{array}{ccc}
\mathbf{1^{\prime \prime}} & \mathbf{1} & \mathbf{1}\\
\mathbf{1} & \mathbf{1^\prime} & \mathbf{1^\prime}\\
\mathbf{1} & \mathbf{1^\prime} & \mathbf{1^\prime}
\end{array}
\right),
& \quad &
\left(
\begin{array}{ccc}
\mathbf{1^\prime} & \mathbf{1^\prime} & \mathbf{1^\prime}\\
\mathbf{1^\prime} & \mathbf{1^\prime} & \mathbf{1^\prime}\\
\mathbf{1^\prime} & \mathbf{1^\prime} & \mathbf{1^\prime}
\end{array}
\right),
\label{LL_1p1pp1pp_1pp1pp1pp}
\end{eqnarray}
respectively.

\section{The fit procedure}
\label{fit}

In this appendix we present the fit procedure adopted in Sec.~\ref{sec:soft}
for the cases in which $A_4$ is softly broken. We define the Hermitian matrices
\begin{eqnarray}
H_\ell
&=&
m_\ell m_\ell^\dagger
=
V_{\ell L}\, \textrm{diag}(m_e^2, m_\mu^2, m_\tau^2)\,  V_{\ell L}^\dagger,
\nonumber\\
H_\nu
&=&
m_\nu m_\nu^\dagger
=
U_\nu^\ast\, \textrm{diag}(m_1^2, m_2^2, m_3^2)\, U_\nu^T.
\end{eqnarray}
With this notation, the PMNS matrix $U$ entering the charged current
interactions as
\begin{equation}
{\cal L}_W = \frac{g}{\sqrt{2}} \bar{\ell}_L^{\textrm{mass}}\,
U\, \gamma^\mu \nu_L^{\textrm{mass}}\, W_\mu^- + \textrm{H.c.},
\end{equation}
is given by $U = V_{\ell L}^\dagger\, U_\nu$.

With a suitable phase choice, we write
\begin{equation}
U = V\, K,
\end{equation}
where the parametrization
\begin{widetext}
\begin{equation}
V=
\left(
\begin{array}{ccc}
c_{12} c_{13} &
s_{12} c_{13} &
s_{13} e^{- i \delta_D}\\
- s_{12} c_{23} - c_{12} s_{23} s_{13} e^{i \delta_D} &
c_{12} c_{23} - s_{12} s_{23} s_{13} e^{i \delta_D} &
s_{23} c_{13}\\
s_{12} s_{23} - c_{12} c_{23} s_{13} e^{i \delta_D} &
- c_{12} s_{23} - s_{12} c_{23} s_{13} e^{i \delta_D} &
c_{23} c_{13}
\end{array}
\right)
\label{V_ckm-pdg}
\end{equation}
\end{widetext}
follows the Particle Data Group notation~\cite{pdg}, and
\begin{equation}
K = \textrm{diag}(1, e^{i \alpha_M/2}, e^{i \beta_M/2})
\end{equation}
contains the Majorana phases. Recent constraints on the mixing angles and
phases of the PMNS matrix can be found in Refs.~\cite{Tortola:2012te,limits}.
The Majorana phases $\alpha_M$ and $\beta_M$ are unconstrained, while
interpretations differ about constraints on the Dirac phase $\delta_D$. We
leave $\delta_D$ free, and follow Ref.~\cite{Tortola:2012te} for the ranges of
the other parameters.

In the cases of interest to us, it is easy to diagonalize $m_\ell$. In the
basis where $m_\ell$ is diagonal, Eqs.~\eqref{diag_mell} and \eqref{diagmnu} hold.
Using the type-I seesaw relation~\eqref{m_nu} and Eq.~\eqref{diagmnu}, we find
\begin{equation}
M_R = - m_D^T\, U\, \textrm{diag}(m_1^{-1}, m_2^{-1}, m_3^{-1})\, U^T\, m_D.
\label{MR_bottomup}
\end{equation}

We perform our fits in the following fashion:
\begin{enumerate}
\item We use random values for $s_{12}$, $s_{13}$, $s_{23}$, $\Delta
    m^2_{21}$, and $\Delta m^2_{31}$, within the 1$\sigma$ intervals found
    in Ref.~\cite{Tortola:2012te}, corresponding to the normal hierarchy.
\item We generate random values for $\delta_D$, $\alpha_M$, $\beta_M$, and
    for $m_1$, keeping the latter between 0 and 0.2~eV.
\item We generate random values for the theoretical parameters in the
    matrix $m_D$, written in the basis where $m_\ell$ is diagonal.
\item We obtain $M_R$ from Eq.~\eqref{MR_bottomup}---by construction,
$m_D$ and $M_R$ are consistent with the experimental observations.
\item We define a figure of merit $\sigma$, which measures the difference
    between the form of $M_R$ obtained from Eq.~\eqref{MR_bottomup} and that
    predicted in Eqs.~\eqref{MR_11p1pp} or \eqref{MR_3} when the symmetry is
    exact.
\item To keep the soft breaking perturbative, only cases where $\sigma$ is
    smaller than some reference value are maintained.
\end{enumerate}

The only cases where soft breaking of $A_4$ allows the type-I seesaw mechanism
to fit the experimental data are cases i), ii), and iii) in
Table~\ref{tabA4:rep}. In the first two cases, $\nu_R \sim \mathbf{3}$, and
$M_R$ has, in the exact $A_4$ limit, the form in Eq.~\eqref{MR_3}. We wish to
keep deviations from this form somewhat small. To be precise, we define
\begin{equation}
\sigma = \sum_{i,j\, = 1}^3 \left| \Sigma_{i j} \right|^2,
\label{sigma1}
\end{equation}
with
\begin{equation}
\Sigma =
\frac{M_R}{\left| (M_R)_{11} \right|}
- \text{diag}\,(1,1,1),
\end{equation}
where $M_R$ comes from Eq.~\eqref{MR_bottomup} in the fit explained above. We
only keep points where $\sigma < 1$.

In case iii), $\nu_R \sim (\mathbf{1}, \mathbf{1^{\prime}}, \mathbf{1^{\prime
\prime}})$, and $M_R$ has the form in Eq.~\eqref{MR_11p1pp} in the exact $A_4$
limit. For this case we define
\begin{equation}
\sigma =
\frac{\left| (M_R)_{12} \right|^2 + \left| (M_R)_{13} \right|^2 +
\left| (M_R)_{22} \right|^2 + \left| (M_R)_{33} \right|^2 }{
\textrm{min}
\left(
\left|(M_R)_{11}\right|^2 , \left|(M_R)_{23}\right|^2\right)
}
\label{sigma2}
\end{equation}
and only keep points where $\sigma < 1$.

\end{document}